\documentclass[journal]{IEEEtran}

\usepackage[american,cute inductors,smartlabels]{circuitikz}
\usepackage{booktabs}	
\usepackage{tabularx}
\usepackage{balance}
\usepackage{cite}
\usepackage{amsfonts,amsmath}
\usepackage{dsfont}
\usepackage{multirow}
\usepackage{graphicx}
\usepackage{lscape}	
\usepackage{url}
\usepackage{siunitx}
\usepackage{tikz}
\usetikzlibrary{arrows,automata}
\usetikzlibrary{arrows,shapes,backgrounds,calc,positioning,patterns}

\DeclareMathOperator{\diag}{diag}
\newcommand{\numberset}{\mathbb}							
\newcommand{\R}{\numberset{R}}

\newcommand{\bR} { {\mathbb R}}
\newtheorem{remark}{Remark}
\newtheorem{proposition}{Proposition}
\newtheorem{lemma}{Lemma}

\newtheorem{assumption}{Assumption}
\newtheorem{objective}{Objective}

\usetikzlibrary{calc}
\ctikzset{bipoles/thickness=1} 
\ctikzset{bipoles/length=1cm}
\tikzstyle{everynode}=[font=\small] \tikzstyle{every path}=[line width=0.8pt,line cap=round,line join=round]

\ifCLASSINFOpdf
  
\else
 
\fi

\begin{document}
%
\title{Distributed control of DC grids:\\ integrating prosumers' motives}
%
%
%

\author{M.~Cucuzzella, 
        T. Bouman, 
        K. C.~Kosaraju,
        G. Schuitema,
        N. H. Lemmen,
        S. Johnson-Zawadzki, 
        C. Fischione, 
        L. Steg,
        J. M. A. Scherpen 
\thanks{M. Cucuzzella is with the Department of Electrical, Computer and Biomedical Engineering, University of Pavia, 27100 Pavia, Italy. He is also with Jan C. Willems Center for Systems and Control, ENTEG, University of Groningen, Nijenborgh 4, 9747 AG Groningen, the Netherlands (email: michele.cucuzzella@unipv.it, m.cucuzzella@rug.nl).}
\thanks{T. Bouman, N. H. Lemmen, S. Johnson-Zawadzki and L. Steg are with the University of Groningen, Grote Kruisstraat 2/1, 9712 TS Groningen, the Netherlands (e-mail: \{t.bouman, n.h.lemmen, s.johnson.zawadzki, e.m.steg\}@rug.nl).}
\thanks{K. C. Kosaraju is with the Department of Electrical Engineering, University of Notre Dame, Notre Dame, IN 46556 USA (e-mail: kkosaraj@nd.edu).}
\thanks{G. Schuitema is with with the University College Dublin, College of Business, Carysfort Avenue, Blackrock, Dublin, Ireland  (e-mail: geertje.schuitema@ucd.ie)}
\thanks{C. Fischione is with the School of Electrical Engineering and Computer Science, KTH Royal Institute of Technology, Stockholm 114 28, Sweden (e-mail: carlofi@kth.se).}
\thanks{J. M. A. Scherpen is with
the Jan C. Willems Center for Systems and Control, ENTEG, University of Groningen, Nijenborgh
4, 9747 AG Groningen, the Netherlands (email: j.m.a.scherpen@rug.nl).}
\thanks{This work is supported by the EU Project ‘MatchIT’ (project number: 82203).}
}



\maketitle

\begin{abstract}
	In this paper, a novel distributed control strategy addressing a (feasible) psycho-social-physical welfare problem in {islanded} Direct Current (DC) smart grids is proposed. Firstly, we formulate a (convex) optimization problem that allows prosumers to share current with each other, taking into account the technical and physical aspects and constraints of the grid (e.g., stability, safety), as well as psycho-social factors (i.e., prosumers' personal values). Secondly, we design a controller whose (unforced) dynamics represent the continuous time primal-dual dynamics of the considered optimization problem. Thirdly, a passive interconnection between the physical grid and the controller is presented. Global asymptotic convergence of the closed-loop system to the desired steady-state is proved and simulations based on collected data on psycho-social aspects illustrate and confirm the theoretical results.
\end{abstract}


%
\IEEEpeerreviewmaketitle

\section{Introduction}
\IEEEPARstart{T}{ransitioning} towards 100\% renewable energy systems
brings about many challenges, requiring solutions
that consider physical, technical as well as psycho-social aspects of energy
systems. Even if renewable energy technologies are able
to deliver the energy demanded, changes in end-users’ energy-related
choices, roles and behaviors are needed to guarantee
the efficiency and sustainability of the energy system \cite{sovacool2014diversity,stern2016opportunities}. 

In the current paper, we focus on achieving proportional current sharing: the situation in which prosumers 
share their generated or stored energy from renewable
sources with peers within their local electricity network, proportionally to the corresponding generation (and/or storage) capacities.
Importantly, given the central role of end-users in renewable
energy systems, we not only look at the commonly considered
technical and physical aspects of energy systems when designing the control scheme, but also
pioneer with integrating psycho-social aspects in our energy control systems, where \lq prosumers' are agents that both produce and consume (or store) energy \cite{prosumer}.
Specifically, we develop a psycho-social model that, based on prosumers' motives (i.e., personal values), predicts the level of flexibility allowed in their local energy grid, which we integrate in our control scheme. We estimate our model coefficients on national representative data (in Section \ref{sec:results}, for the Netherlands), enabling us to make predictions about the level of flexibility in specific subpopulations (e.g., local energy districts).
Through this approach, we provide first insights in,
and promote and facilitate, the so needed integration
of more psycho-social factors in the modeling and optimization of
energy systems \cite{sovacool2014diversity,stern2016opportunities}.
Thereby, we not only provide insights in what is technically possible, but also in what is socially feasible.

From a technical perspective, although inverter-based Alternate Current (AC) microgrids have been prevalent, the recent wide spread of
renewable energy sources, electronic appliances and batteries (including for instance also electric vehicles) motivates the design and operation of islanded DC smart grids \cite{hatziargyriou2014microgrids,Nasirian2015, 8468058,8444703,8556484,8587203,cucuzzella2017robust}, which are generally simpler to control and more efficient and reliable than AC networks \cite{Justo2013387,Nasirian2015}, so contributing to the transition towards a 100\% renewable energy system. For example, lossy DC-AC and AC-DC conversion stages are reduced, frequency and reactive power control is not necessary, frequency synchronization issues are overcome, harmonics and skin effect are absent. To guarantee a proper and safe
functioning of the power network, voltage stabilization is the
main goal to achieve in DC smart grids \cite{7268934}. Additionally, to
avoid the overstressing of a single energy source or storage unit, it is generally
desired that the total demand is proportionally shared among
all the prosumers of the smart grid \cite{Nasirian2015}. 
These objectives play a key role specially in islanded grids that mainly rely on renewable energy. Furthermore, in order to move towards a 100\% renewable energy system and due to the absence of the main grid and conventional fuel-based generators, it is common to adopt demand response programs \cite{SIANO2014461}, that allow automation to control prosumers' loads, maintaining voltage stability and high power quality.
However, to permit
prosumers to share their generated current or power, voltage
differences among the nodes of the smart grid are necessary.

In the literature, several control techniques have been proposed
to control the voltages towards the corresponding nominal
values (see for instance \cite{7983406,Ferguson,Silani} and the references therein).
Other works have proposed consensus-based
control schemes achieving current/power sharing without
regulating the voltage (see for instance \cite{DePersis2016} and the references
therein). Differently from the above mentioned works,
consensus-based protocols have been recently designed for
achieving both current sharing and a particular form of voltage
regulation, where the average value of the voltages of the
whole microgrid is controlled towards a desired setpoint (see
for instance \cite{Nasirian2015,cucuzzella2017robust,J_Trip2018} and the references therein).
However, regulating \emph{only} the average voltage may introduce, in
some nodes of the microgrid, large voltage deviations from the
corresponding nominal value, making this solution not always
adequate in practical applications. This motivated us to design an optimal
control scheme aiming to share, among the prosumers of the
smart grid, the largest possible amount of the (controllable) total
demand in compliance with physical constraints that ensure
safety and reliability (a similar control problem is addressed
in \cite{8796095} by synthesizing a centralized symbolic controller), while considering psycho-social aspects of the involved prosumers as well.

{Critically, like other energy solutions \cite{sovacool2014diversity,stern2016opportunities,Dietz18452}, the performance of the proposed control system strongly depends on the number of prosumers that is willing to adopt it, and the degree to which these prosumers allow the optimal control scheme to control their load demand (i.e., flexibility). Such decisions of prosumers are influenced by many psycho-social factors, of which many are rooted in individuals' personal values, reflecting general, overarching motivations that guide individuals' beliefs and actions \cite{Steg2015,VANDERWERFF20158,steg2018drives,Dietz14081,stern1994value,BOUMAN2020102061,BOUMAN2020b}, and which are focal in the current paper. Two general clusters of values typically motivate individuals' energy behaviours: self-transcendence values (STV), reflecting personal goals to care for nature and the environment, other people and social welfare; and self-enhancement values (SEV), reflecting personal goals to acquire possessions, status, pleasure and comfort \cite{Stern_Brief,WESLEYSCHULTZ1999255,Schwartz1992}.} 

Whereas all individuals endorse both types of values to some extent, they differ in how much they endorse and prioritize each of these values. These levels of endorsement and prioritizations robustly guide and predict individuals' beliefs and actions. Specifically, individuals are more likely to engage in actions that support strongly endorsed and prioritized values, and are more likely to refrain from actions that threaten strongly endorsed and prioritized values \cite{Schwartz1992,Schwartz2012}. Applied to the context of this paper, prosumers may for instance adopt an automated control scheme because they see it as serving sustainability (i.e., self-transcending) or financial (i.e., self-enhancement) goals, and/or because they see it as serving a convenience goal of not having to control appliances oneself (i.e., self-enhancement). Conversely, if automated control means that appliances are operated at other than desired ways or at inconvenient times, high levels of automated control may challenge self-enhancement values, which may refrain prosumers from adopting high levels of flexibility, particularly if they strongly endorse self-enhancement values. We will incorporate this logic in the design and tuning of the proposed control scheme (see Subsection \ref{subsec:social}). We think that this first integration of prosumers' motives, and at later stages other psycho-social factors, in the modelling and design of energy control systems could greatly contribute to a transition towards a 100\% renewable energy system, enabling the more realistic prediction and assessment of renewable energy systems feasibility. Moreover, it can provide valuable insights in how flexibility could be made more acceptable, desirable and realistic, for instance by ensuring flexibility benefits strongly endorsed, and highly influential personal values.

More concretely, in this paper, we consider an {islanded} DC smart grid with a number of prosumers interconnected through resistive-inductive transmission lines. For the considered DC smart grid we propose a novel distributed optimal control scheme that addresses a psycho-social-physical welfare problem and allows to share among the prosumers financial, technical and social costs and utilities associated with the generation and consumption of energy, fulfilling (at the steady-state) physical requirements.

To achieve these goals we use an approach that bridges convex optimization and systems theory, i.e., (continuous) primal-dual dynamics \cite{kose1956solutions,feijer2010stability,cherukuri2016asymptotic} and passivity \cite{van2000l2}. 
The contributions of the paper can be summarized as follows:  
\begin{itemize}
	\item[(1)] {We formulate a (convex) psycho-social-physical welfare optimization problem.}
	\item[(2)] We design a distributed optimal controller, whose (unforced) dynamics represent the primal-dual dynamics of the considered optimization problem.
	\item[(3)] After showing the passivity properties of the smart grid and the controller, a power-conserving interconnection between the smart grid and the controller is established and the (global) asymptotic convergence of the closed-loop system trajectories to the desired equilibrium point is proved.
	\item[(4)] Based on the personal values of prosumers, we provide $i$) a procedure to estimate the acceptable level of automation (i.e., flexibility) within a given local energy grid and $ii$) a simple rule to tune some controller parameters.
\end{itemize}

The remainder of this paper is outlined as follows. In Section~\ref{sec:model} we describe the model of the considered DC smart grid and show its passivity property. Thereafter, in Section~\ref{sec:pf} we state the desired control objectives and integrate prosumers' motives. In Section~\ref{sec:controller} we explain the proposed control strategy and provide stability analysis. In Section~\ref{sec:results}, we asses the proposed control scheme in simulation based on collected data on psycho-social aspects from the Netherlands and conclude with some {future research directions} in Section~\ref{sec:conclusions}.

{\bf Notation}. The set of real numbers and non-negative real numbers are denoted by $\bR$ and $\bR_+$, respectively. For a vector $x \in \bR^n$ and a symmetric and positive semidefinite matrix~$M \in \bR^{n \times n}$, let~$\| x\|_M:= (x^\top M x)^{1/2}$. If~$M$ is the identity matrix, this is the Euclidean norm and is denoted by~$\|x\|$. For symmetric matrices $P,Q \in \bR^{n\times n}$, $P \le Q$ implies that $Q-P$ is positive semidefinite. $\mathds{I}$ and $\mathds{1}$ denote the identity matrix and ones vector of appropriate dimensions, respectively, while ${\bf0}$ denote the null matrix (or vector) of appropriate dimensions. Let $x, u$ be the state and input of the physical plant, $x^\ast$ and $u^\ast$ denote the corresponding optimization variables. { Moreover, $(\overline{u},\overline{x})$  and $(\overline{u}^\ast, \overline{x}^\ast)$ denote the values of $(u,x)$ and $(u^\ast,x^\ast)$  at the steady-state, respectively.}
%
%
%
%
\section{DC Smart Grid}
\label{sec:model}
 \begin{figure}[t]
    	\begin{center}
    		\begin{circuitikz}[scale=.95,transform shape]
    			\ctikzset{current/distance=1}
    			\draw
    			node[] (Ti) at (0,0) {}
    			node[] (Tj) at ($(5.4,0)$) {}
    			node[] (Aibattery) at ([xshift=-4.5cm,yshift=0.9cm]Ti) {}
    			node[] (Bibattery) at ([xshift=-4.5cm,yshift=-0.9cm]Ti) {}
    			node[] (Ai) at ($(Aibattery)+(0,0.2)$) {}
    			node[] (Bi) at ($(Bibattery)+(0,-0.2)$) {}
    			(Ai) to [R, l={$R_{si}$}] ($(Ai)+(1.7,0)$) {}
    			($(Ai)+(1.7,0)$) to [short,i={$I_{si}$}]($(Ai)+(1.701,0)$){}
    			($(Ai)+(1.701,0)$) to [L, l={$L_{si}$}] ($(Ai)+(3,0)$){}
    			to [short, l={}]($(Ti)+(0,1.1)$){}
    			(Bi) to [short] ($(Ti)+(0,-1.1)$);
    			\draw
    			($(Ai)$) to []($(Aibattery)+(0,0)$)to [V=$u_{si}$]($(Bi)$)
    			($(Ti)+(-1.3,1.1)$) node[anchor=south]{{$V_{i}$}}
    			($(Ti)+(-1.3,1.1)$) node[ocirc](PCCi){}
    			($(Ti)+(-.3,1.1)$) to [short,i>={$I_{li}u_{li}$}]($(Ti)+(-.3,0.5)$)to [I]($(Ti)+(-.3,-1.1)$)
    			($(Ti)+(-1.3,1.1)$) to [C, l_={$C_{i}$}] ($(Ti)+(-1.3,-1.1)$)
    			($(Ti)+(2.,1.1)$) to [short,i={$I_{ij}$}] ($(Ti)+(2.2,1.1)$)
    			($(Ti)+(0,1.1)$)--($(Ti)+(.6,1.1)$) to [R, l={$R_{ij}$}] 
    			($(Ti)+(2.5,1.1)$) {} to [L, l={{$L_{ij}$}}, color=black]($(Tj)+(-2.2,1.1)$){}
    			($(Tj)+(-2.2,1.1)$) to [short]  ($(Ti)+(3.4,1.1)$)
    			($(Ti)+(0,-1.1)$) to [short] ($(Ti)+(3.4,-1.1)$);
    			\draw
    			node [rectangle,draw=none,minimum width=5.9cm,minimum height=3.6cm,dashed,fill=cyan,opacity=0.1,label=\textbf{Prosumer (Pr) $i$},densely dashed, rounded corners] (DGUi) at ($0.5*(Aibattery)+0.5*(Bibattery)+(2.5,0.4)$) {}
			node [rectangle,draw=none,minimum width=1.5cm,minimum height=2.6cm,fill=red,opacity=0.2,dashed,label=\textbf{Load $i$},densely dashed, rounded corners] (DGUi) at ($0.5*(Aibattery)+0.5*(Bibattery)+(4.5,0)$) {}
			node [rectangle,draw=none,minimum width=2.25cm,minimum height=3.6cm,fill=gray,opacity=0.3,label=\textbf{Line $ij$}, rounded corners] (DGUi) at ($0.5*(Aibattery)+0.5*(Bibattery)+(6.65,0.4)$) {};
    		\end{circuitikz}
    		\caption{Electrical scheme of prosumer $i \in \mathcal{V}$ and transmission line $k \sim \{i, j\} \in \mathcal{E}$, $j \in \mathcal{N}_i$, where $\mathcal{N}_i$ is the set of the prosumers connected to prosumer $i$.}
    		\label{fig:microgrid}
    	\end{center}
    	\vspace{.2cm}
    \end{figure}
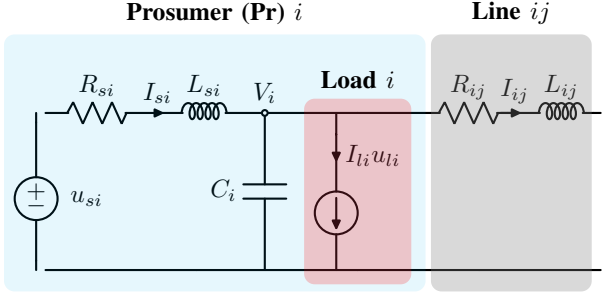
    
 	In this paper we consider a low-voltage {islanded} DC smart grid with $n$ prosumers connected to each other through $m$ resistive-inductive ($RL$) transmission lines {\cite{Nasirian2015,8556484,8468058,7268934,7983406,DePersis2016,cucuzzella2017robust,J_Trip2018,8796095}}. 
	For the readers' convenience, a schematic electrical diagram of the considered  smart grid is illustrated in Fig.~\ref{fig:microgrid} (see also Table~\ref{tab:symbols} for the description of the used symbols).
	Each prosumer is represented by a DC voltage source\footnote{With a slight abuse of nomenclature, the considered DC voltage source can also represent for instance the output voltage of an energy storage system, whose sizing  is out of the scope of the paper. Yet, thanks to the optimal control scheme we design in the next sections, which can reduce the energy consumption of some electrical appliances of those prosumers who adopt automation, we think that the costs associated to the energy storage systems will not be high.} supplying a controllable load $I_{li}u_{li}$. More precisely, given the load demand $I_{li}$ of prosumer $i$, the control input $0\leq u_{li}^{\min}\leq u_{li} \leq 1$ can reduce the demand to $I_{li} u_{li}^{\min}$, which represents the current required to supply the {uncontrollable} loads of prosumer $i$. 
	{Based on technical data, we will provide a procedure to estimate in Subsection \ref{subsec:inequalities1} the parameter $u_{li}^{\min}$, corresponding to the maximum amount of flexibility that would be technically possible if prosumer $i$ accepts to adopt automation of different appliances (see also Section \ref{sec:results} where we use real data). Specifically, we observe that the term $1-u_{li}^{\min}$ represents the proportion of electricity consumed within a household that could potentially be automated from a technical point of view if prosumer $i$ accepts to adopt automation. However, not all prosumers will accept flexibility, and not all prosumers will accept flexibility to the same degree (e.g., some might allow all possible appliances to be flexible, whereas others only allow some appliances to flexible). In Subsection \ref{subsec:social} we provide a procedure to  estimate the acceptable level of automation (i.e., flexibility) within a smart grid.} 	
	\begin{table}
		\begin{center}
			\caption{Description of the used symbols}\label{tab:symbols}
			\begin{tabular}{cl}
				\toprule
				{\bf Symbol} &{\bf Description}\\
				\midrule
				$I_{si}$						& Generated current\\
				$R_{si}, L_{si}$ & Filter resistance, inductance\\
				$V_i$						& Load voltage\\
				$C_{i}$ &Shunt capacitor\\
				$I_{k}$ 						& Line current\\
				$R_k, L_k$ & Line resistance, inductance\\
				$u_{si}, u_{li}$						& Control inputs\\
				$I_{li}$ & Load current\\
				\bottomrule
			\end{tabular}
		\end{center}
	\end{table}
	
	The overall DC smart grid is represented by a connected and undirected graph $\mathcal{G} = (\mathcal{V},\mathcal{E})$, where the nodes, $\mathcal{V} = \{1,...,n\}$, represent the prosumers and the edges, $\mathcal{E}  = \{1,...,m\}$, represent the transmission lines interconnecting the prosumers.
	Therefore, the topology of the smart grid is described by its corresponding incidence matrix $\mathcal{B} \in \R^{n \times m}$. The ends of edge $k \in \mathcal{E}$ are arbitrarily labeled with a $+$ and a $-$, and the entries of  $\mathcal{B}$ are given by $\mathcal{B}_{ik}=+1$ if $i$ is the positive end of $k$, $\mathcal{B}_{ik}=-1$ if $i$ is the negative end of $k$, and $\mathcal{B}_{ik}=0$ otherwise.
	Consequently, the overall dynamical system describing the smart grid behavior can be written compactly for all the prosumers $i \in \mathcal{V}$ as
\begin{subequations}\label{eq:plant}
\begin{align}
		L_s\dot{I}_{s} & =-R_s I_s- V + u_s\label{p1}\\
		L\dot{I} & = - R I -\mathcal{B}^\top V\label{p2}\\
		C\dot{V} & = I_{s} + \mathcal{B}I - I_l u_l\label{p3},		
		\end{align}
		\end{subequations}
	where $I_s,V, u_s, u_l: \R_+ \to  \R^{n}$ and {$I: \R_+\to \R^m$}.
	Moreover, $C, L_s, R_s, I_l \in \R^{n \times n}$ and $R, {L} \in \R^{m \times m}$ are positive definite diagonal matrices, e.g., $I_l = \diag(I_{l1}, \dots, I_{ln})$. 
	Furthermore, let $x:=[I_s^\top,I^\top,V^\top]^\top \in \mathcal{X} \subseteq \R^{2n+m}$ and $u := [u_s^\top, u_l^\top]^\top \in \mathcal{U} \subseteq \R^{2n}$ denote the state and input of system \eqref{eq:plant}, respectively.
	Then, for a given constant input $\overline u$, the corresponding steady state solution $(\overline I_s,\overline I,\overline V)$ to system~\eqref{eq:plant} satisfies 
	\begin{subequations}\label{eq:steady_state_solution2}
		\begin{align}
		\overline{V} & =  -R_s\bar{I}_s+\overline{u}_s \label{ss1}\\
		\overline I & = -R^{-1}\mathcal{B}^\top\overline V \label{ss3}\\
		\overline{I}_{s} &=  - \mathcal{B} \overline I + I_l \overline{u}_l. \label{ss2}
		\end{align}
	\end{subequations}
Before establishing a useful property of system \eqref{eq:plant}, 
we first define the set of all feasible forced equilibria of \eqref{eq:plant} as follows:
		\begin{align}\label{set:equilibrium}
E :=\{ (\overline{u},\overline{x})\in \mathcal{U}\times\mathcal{X}| (\overline{u},\overline{x}) ~ \text{satisfies} ~ \eqref{eq:steady_state_solution2}\}.
\end{align}
	Moreover, we observe that for any $u=\overline{u}$, the steady-state solution $\overline{x}$ to \eqref{eq:plant} is unique and satisfies
\begin{subequations}	\label{eq:unique_ss}	
	\begin{align}
		\overline{V}&= (\mathds{I}+R_s \mathcal{L})^{-1}\left(\overline{u}_s-R_s I_l \overline{u}_l\right)\label{uss1}\\
				\overline{I}_s&= \mathcal{L}(\mathds{I}+R_s \mathcal{L})^{-1}\left(\overline{u}_s-R_s I_l\overline{u}_l\right)+I_l\overline{u}_l\label{uss2}\\
						\overline{I}&= -R^{-1}\mathcal{B}^{\top}(\mathds{I}+R_s \mathcal{L})^{-1}\left(\overline{u}_s-R_s I_l\overline{u}_l\right),\label{uss3}
	\end{align}
\end{subequations}
where $\mathcal{L}:=\mathcal{B} R^{-1}\mathcal{B}^\top$ is the (weighted) Laplacian matrix associated with the physical network.

	Now, in analogy with \cite{2018arXiv181102838C}, we establish a passivity property of system \eqref{eq:plant} that will be useful in Section \ref{sec:controller} for the controller design.
	
	\begin{proposition}{\bf(Passivity property of \eqref{eq:plant})}.\label{prop:microgrid} 
		Let $y:=[\dot{I}_s^\top, -\dot{V}^\top I_l]^\top$ and $u_d := [u_{sd}^\top, u_{ld}^\top]^\top$, $u_{sd},u_{ld}:\mathbb{R}_+\to \mathbb{R}^{n}$. The following statements hold:
		\begin{itemize}
			\item[(a)] System \eqref{eq:plant} together with $\dot{u}=u_d$ is passive with respect to the supply rate $u_d^\top y$ and storage function
			\begin{equation}\label{eq:Storage_fun}
				S(u,x)=\dfrac{1}{2} \dot{x}^\top M \dot{x},
			\end{equation}
with $M:=\diag\{L_s, L, C\}$.
			\item[(b)]Let $u_d={\bf0}$. The state of system \eqref{eq:plant} converges to the equilibrium point $(\overline{u},\overline{x}) \in E$.
		\end{itemize}
	\end{proposition}
\begin{IEEEproof}
The storage function $S$ in \eqref{eq:Storage_fun} satisfies
\begin{align}
\label{eq:dS}
\begin{split}
\dot{S} &= -\dot{I}_s^\top R_s \dot{I}_s -\dot{I}^\top R \dot{I} +u_{sd}^\top\dot{I}_s -u_{ld}^\top I_l \dot{V}\\
&\leq u_{sd}^\top\dot{I}_s -u_{ld}^\top I_l \dot{V},
\end{split}
\end{align}
along the solutions to \eqref{eq:plant}, concluding the proof of part (a).
For part (b), we conclude from \eqref{eq:dS} that there exists a forward invariant set $\Omega$ and by LaSalle's invariance principle the solutions that start in $\Omega$ converge to the largest invariant set contained in
\begin{equation}
\Omega \cap \left\{({u},{x})\in \mathcal{U}\times\mathcal{X}|\dot{u}={\bf0}, \dot{I}_s={\bf0},\dot{I}={\bf0} \right\}.
\end{equation}
Moreover, from the first line of \eqref{eq:plant} it follows that $V$ is also a constant vector in $\Omega$. Then, the solutions that start in $\Omega$ converge to the largest invariant set contained in $\Omega \cap {E}$, concluding the proof of part (b).
\end{IEEEproof}
%
%
%
%
\section{Problem Formulation}
\label{sec:pf}
In this section, 
 we formulate and discuss the main goal of the paper, which allows to share among the prosumers of the smart grid the financial, technical and social costs and utilities associated with the generation and consumption of energy.

Firstly, we notice that \eqref{eq:steady_state_solution2} implies the following equalities:  
	\begin{subequations}\label{eq:steady_state_solution_reduced}
		\begin{align} 
		{\bf0} & = \overline{u}_s - R_s\bar{I}_s-\overline{V},\label{ssr1}\\
		{\bf0} &=  - I_l\overline{u}_l + \overline{I}_{s} - \mathcal{L}\overline{V}.\label{ssr2}
		\end{align}
	\end{subequations}

Now, let $x_r:=[I_s^\top,V^\top]^\top \in \mathcal{X}_r \subseteq \R^{2n}$ denote the reduced state of system \eqref{eq:plant}. Then, let us define the following set\footnote{Note that $(\overline{u}, \overline{I}_s, \overline V) \in {E}_r \iff (\overline{u},\overline{I}_s,- R^{-1}\mathcal{B}^\top\overline V,\overline V) \in {E}$.}:
\begin{align}\label{set:equilibrium_a}
E_r :=\{ (\overline{u},\overline{x}_r) \in \mathcal{U}\times\mathcal{X}_r |(\overline{u}, \overline{x}_r)~ \text{satisfies}~ \eqref{eq:steady_state_solution_reduced}\},
\end{align}
which we will use as set of equality constraints of the optimization problem we formulate later in this section.
\subsection{Prosumer's cost and utility}
\label{subsec:C_U}
We observe that the second line of \eqref{eq:steady_state_solution_reduced} implies that at the steady-state the total generated current $\mathds{1}^\top \overline{I}_s$ is equal to the total current demand $\mathds{1}^\top I_l\overline{u}_l$.
Therefore, there is flexibility to distribute the total required current optimally among the various (equivalent) prosumers.
Generally, in order to achieve an efficient demand and supply matching, so avoiding the overstressing of a source, it is desirable  that the total load demand of the smart grid is shared among all the prosumers proportionally to the corresponding generation (and/or storage) capacities (\emph{proportional} current sharing). This desire is equivalent to achieving ${\overline{I}_{si}}/{ {\pi}_{ci}} ={\overline{I}_{sj}/ {\pi}_{cj}}$ for all $i,j \in \mathcal{V}$, where a relatively large value of $ {\pi}_{ci}\in \R_+$ corresponds for instance to a relatively large generation (and/or storage) capacity of prosumer~$i$. We call this desire \emph{ideal} current sharing and, in analogy with \cite{cucuzzella2017robust}, can be expressed as follows:
\begin{align}
\label{eq:current_sharing}
\lim_{t \rightarrow \infty} I_s(t)= \overline{I}_{s} =   {\Pi}_c\mathds{1}{i}_{s},
\end{align}
with $ {\Pi}_c = \diag( {\pi}_{c1},\dots, {\pi}_{cn})$ and ${i}_{s}=\mathds{1}^\top I_l \overline{u}_l/({\mathds{1}^\top  {\Pi}_c\mathds{1}})$. Moreover, without loss of generality, we assume that $\sum_{i=1}^n \pi_{ci} = 1$, i.e., given a certain generation or storage capacity of node~$i$ (say $I_{si}^\mathrm{base}$), then $\pi_{ci}=I_{si}^\mathrm{base}/\sum_{i=1}^n I_{si}^\mathrm{base}$.

A transition towards 100\% renewable energy systems requires that end-users change their energy-related behaviours and accept new technologies such as \emph{demand response} \cite{SIANO2014461} programs, which control the prosumers' appliances.
Therefore, to make the notion of optimality explicit, we assign to every prosumer $i$ a strictly convex quadratic \lq cost' function $C_i(I_{si})$ related to the generated current $I_{si}$ and a strictly concave quadratic \lq utility' function $U_i(u_{li})$ related to the current consumption $I_{li}u_{li}$, i.e., 
\begin{subequations}\label{eq:cost_utility}
		\begin{align}
		C_i(I_{si}) &=  \frac{I_{si}^2}{2  {\pi}_{ci}}  \quad \forall i \in \mathcal{V}, \label{cost}\\
		U_i(u_{li}) &=  -\frac{I_{li}^2 \left(1-u_{li}\right)^2}{2  {\pi}_{ui}}  \quad \forall i \in \mathcal{V}, \label{utility}
		\end{align}
	\end{subequations}
	where a relatively small value of $ {\pi}_{ui} \in \R_+$ corresponds for instance to a relatively large request of comfort from prosumer $i$.
	{While $\pi_{ci}$ has the physical interpretation explained above, we provide in Subsection \ref{subsec:social} a motives-based interpretation (and choice) of the coefficient ${\pi}_{ui}$ appearing in the utility function.}
	
\subsection{Psycho-social-physical welfare}
	
Let $C(I_s):=\sum_{i \in \mathcal{V}} C_i(I_{si})$ and $U(u_l):=\sum_{i \in \mathcal{V}} U_i(u_{li})$. Then, we denote the \emph{psycho-social welfare} by $W(u_l,I_s):=U(u_l)-C(I_s)$ and consider the following convex minimization problem:
\begin{align}\label{eq:min1}
		\begin{split}
		&\underset{u^\ast, x_r^\ast} {\min} -W\left(u_l^\ast, I_s^\ast \right) \\
		&\text{s.t.} \left( \overline{u}^\ast,\overline{x}_r^\ast \right)\in {E_r}. 
		\end{split}
	\end{align}
Considering the Lagrangian function associated with the optimization problem \eqref{eq:min1} and manipulating the first-order optimality conditions leads to the following lemma, which makes the solution to \eqref{eq:min1} explicit:
\begin{lemma}{\bf (Optimal psycho-social welfare)}.
\label{lemma:min1}
The solution to \eqref{eq:min1} satisfies
\begin{subequations}\label{eq:min1_sol}
		\begin{align}
		&\overline{I}_s^\ast =   {\Pi}_c \mathds{1} \lambda^\mathrm{opt} \label{mina_sol}\\
		&I_l\overline{u}_l^\ast = \left( I_l - \lambda^\mathrm{opt} {\Pi}_u\right)\mathds{1} \label{minb_sol},
		\end{align}
	\end{subequations}
			where
		\begin{equation}
		\label{eq:lambda_opt}
		\lambda^\mathrm{opt} = \frac{\mathds{1}^\top I_l\mathds{1}}{\mathds{1}^\top \left(  {\Pi}_c +  {\Pi}_u \right)\mathds{1}},
		\end{equation}
		with $ {\Pi}_u = \diag( {\pi}_{u1},\dots, {\pi}_{un})$.
\end{lemma}	
Moreover, note that \eqref{mina_sol} implies indeed that \emph{ideal} current sharing is achieved (see \eqref{eq:current_sharing}).
{Furthermore, with the choice $\sum_{i=1}^n \pi_{ci} = 1$, we obtain from \eqref{minb_sol} and \eqref{eq:lambda_opt}  that at the steady-state the consumption reduction is given by
\begin{equation}
\label{eq:saved_current}
I_l(\mathds{1}-\overline{u}_l^\ast) = \frac{\mathds{1}^\top I_l\mathds{1}}{1 + \mathds{1}^\top {\Pi}_u \mathds{1}}\Pi_u\mathds{1},
\end{equation}
which implies that the degree to which prosumers allows the optimal control scheme to control their own load demand strongly depends on $\Pi_u$, where relatively large values of $\pi_{ui}$ correspond to prosumers allowing the control system to control a relatively large proportion of their consumption.}

Now, we assume that at the PCC of each prosumer $i \in \mathcal{V}$, there exists a desired voltage:
\begin{assumption}{\bf (Desired voltages)}.
	\label{ass:desired_voltages}
	There exists a desired voltage $V_{di}>0$ for each $i \in \mathcal{V}$.
\end{assumption}
However, achieving \emph{ideal} current sharing, prescribes the value of the required differences in voltages among the nodes of the smart grid. As a consequence, it is generally not possible to control the voltage at each node towards the corresponding desired value. 
For this reason, the voltage requirements are generally relaxed and, as an alternative, several control approaches in the literature propose to regulate the \emph{average} voltage across the whole microgrid towards a global voltage set point~\cite{Nasirian2015,cucuzzella2017robust}, where the sources with the largest generation capacity determine the grid voltage, i.e.,
\begin{align}
\label{eq:voltage_balancing}
\lim_{t \rightarrow \infty} {\mathds{1}^\top  {\Pi}_c V(t)} =  {\mathds{1}^\top  {\Pi}_c \overline{V}} = {\mathds{1}^\top  {\Pi}_c V_d}.
\end{align}
However, we note that achieving \emph{ideal} current sharing even preserving the average voltage of the smart grid may not always be desired, as it may introduce, in some nodes of the microgrid, large voltage deviations from the corresponding desired value.
 Consider for instance a DC smart grid with 2 prosumers interconnected through a purely resistive transmission line, the value of which is relatively large (e.g., because the prosumers are spatially  distant). Moreover, assume that the load demand of one of the prosumers is much higher than the other. Then, in order to achieve \emph{ideal} current sharing, the prosumers need to share a relatively large current through the transmission line, implying a relatively large voltage deviation (with respect to the desired value) at the corresponding PCCs.
 Consequently, a steady-state solution satisfying \eqref{eq:current_sharing} and \eqref{eq:voltage_balancing} may be not \emph{feasible} in practical applications. Therefore, in order to address this physical issue we modify \eqref{eq:min1} as follows:
 \smallskip 
\begin{objective}{\bf (Psycho-social-physical welfare)}.
\label{obj:1}
\begin{subequations}\label{eq:min2}
		\begin{align}
		&\underset{u^\ast, x_r^\ast} {\min} -\alpha W\left(u_l^\ast, I_s^\ast \right) +\frac{\beta}{2}\|u_s^\ast\|^2 +\frac{\gamma}{2}\|V^\ast-V_d\|^2 \label{min2a}\\
		&\text{s.t.} \left( \overline{u}^\ast,\overline{x}_r^\ast \right)\in {E_r}, \label{min2b}
		\end{align}
	\end{subequations}
		where $\alpha,\beta,\gamma \in \mathbb{R}_{+}$ are design parameters.
\end{objective}
\bigskip

\begin{remark}{{\bf (Rationale behind Objective \ref{obj:1})}}.
\label{rm:rationale}
The quadratic function in \eqref{min2a} comprises three different terms concerning { (i)} the psycho-social welfare, {(ii)} the control effort and {(iii)} the voltage deviation from the corresponding desired value.
As a consequence, a solution to Objective 1, generally differs from the solution to \eqref{eq:min1} and does not guarantee the achievement of \emph{ideal} current sharing \eqref{eq:current_sharing}. This leads to a compromise between the psycho-social welfare and physical requirements. In order to ensure a proper and safe functioning of the smart grid, the voltage requirement has a priority higher than current sharing. In other words, we are interested in a \emph{feasible} solution that permits to share among the prosumers of the smart grid the largest possible amount of total (controllable) demand in compliance with physical requirements.
\end{remark}

\subsection{Additional constraints}
\label{subsec:inequalities1}
In this subsection, we introduce a set of additional inequality constraints, which ensure a safer (steady-state) functioning of the prosumers' appliances and {allow to select the level of flexibility that is technically feasible.} 
More precisely, in order to guarantee a proper functioning of the prosumers' appliances, it is generally required that the voltages remain within prescribed limits (see for instance \cite{4448761} and the references therein). 
Therefore, we consider in this paper the following steady-state constraints:
\begin{objective}{{\bf (Voltage constraints)}}.\label{obj:source}
\begin{align}\label{eq:us_limits}
V_{i}^{\min} \leq \lim_{t \rightarrow \infty} {V_{i}(t)}\leq V_{i}^{\max},
\end{align}
where $V_{i}^{\min}, V_{i}^{\max} \in \R_+$ denote the minimum and maximum permitted value of the voltage $V_{i}$, for all $i \in \mathcal{V}$. Moreover, following the rationale outlined in Remark \ref{rm:rationale}, the interval $[V_{i}^{\min}, V_{i}^{\max}]$ represents the $i$-th available voltage band to share the largest possible amount of total (controllable) demand.
\end{objective}

\begin{objective}{{\bf (Load constraints)}}.\label{obj:load}
\begin{align}\label{eq:ul_limits}
u_{li}^{\min} \leq \lim_{t \rightarrow \infty} {u_{li}(t)}\leq 1,
\end{align}
where $u_{li}^{\min} \in [0,1)$ denotes the minimum permitted value of $u_{li}$, for all $i \in \mathcal{V}$. 
\end{objective}

Specifically, if $u_{li}^{\min}$ in \eqref{eq:ul_limits} is close to 0, then the maximum amount of flexibility that would be technically possible is high. Moreover, let $\mathcal{A}=\{1,\dots,a\}$ represent the set of all the electrical appliances within an average household that could potentially be automated.
Then, let $\Psi$ denote the maximum amount of flexibility (normalized with respect to the total consumption $\mathds{1}^\top I_l \mathds{1}$) from a technical perspective, representing the proportion of electricity consumed by the $a$ electrical appliances that could potentially be automated. 
Thus, the proportion $\Psi$ gives an estimation of what levels of flexibility would be technically possible (i.e., if prosumer $i\in\mathcal{V}$ accepts to adopt automation of the $a$ electrical appliances, then $u_{li}^{\min}$ in \eqref{eq:ul_limits} is equal to $1 - \Psi$). Yet, it is unlikely that all individuals will adopt automation, and not all individuals will adopt automation to the same degree, leading to different values of the coefficients $\pi_{ui}$ in the utility function \eqref{utility}, on which we elaborate in Subsection \ref{subsec:social}.


\subsection{A motives-based interpretation}
\label{subsec:social}
In this subsection, inspired by \cite{schwartz1992universals,steg2018drives,stern1994value,bouman2018measuring,steg2014significance}, we provide a motives-based procedure to estimate the acceptable level of automation (i.e., flexibility) within a given local power grid. Then, based on such a level, we provide a simple rule for tuning the coefficients $ {\pi}_{ui}$ appearing in the utility function \eqref{utility} of prosumer $i \in \mathcal{V}$.

As introduced earlier, the level of flexibility that individuals are willing to adopt likely depends on their general motives, here represented by their personal values. It has for instance been suggested that flexibility may serve individuals' self-transcendence values (e.g., by reducing CO2 emissions, by using electricity efficiently) and self-enhancement values (e.g., by saving costs, by taking over the effort of controlling appliances), and thus that individuals with stronger self-transcendence and self-enhancement values are more likely to adopt flexibility. Conversely, it has also been argued that flexibility may sometimes be perceived as having costs for self-enhancement values (e.g., not being able to use appliances at one's own convenience), and thus that individuals with stronger self-enhancement values are less likely to adopt flexibility. Accordingly, we argue that the level of flexibility in a specific population can be estimated from the following linear model:
\begin{equation} 
\label{eq:rho_j}
\rho_j = \mu_j + \theta_j \; \mathrm{STV} + \epsilon_j \; \mathrm{SEV}, 
\end{equation}
where $\mu_j$ represents the proportion of adopted flexibility in a population (e.g., at national level), STV represents the endorsement of self-transcendence values and SEV the endorsement of self-enhancement values in the focal subpopulation (e.g., a local energy community) whose energy system will be optimized. Depending on the coefficients $\theta_j$ and $\epsilon_j$ respectively, the levels on STV and SEV will either result in lower or higher estimations of flexibility for the subpopulation $\rho_j$, compared to what is observed in the overall population $\mu_j$.

We suggest to estimate the level of flexibility in local energy systems based on national data ($\mu_j, \theta_j, \epsilon_j$) and (assumptions about) the values being endorsed within targeted subpopulation(s) (i.e., STV and SEV in \eqref{eq:rho_j}), on which relatively rich data is available as such values are commonly measured across diverse samples, and for many different research purposes (e.g., European Social Survey\footnote{{https://www.europeansocialsurvey.org/}}).

Now, let $\Lambda$ denote the acceptable level of flexibility in the considered smart grid for the $a$ electrical appliances. Then, we obtain
\begin{equation}
\label{eq:Lambda}
\Lambda = \Psi \sum_{j\in\mathcal{A}}\omega_j \rho_j, 
\end{equation}
where the weight $\omega_j$ reflects how much the appliance $j\in\mathcal{A}$ contributes to the households' energy consumption, satisfying $\sum_{j\in\mathcal{A}} \omega_{j} = 1$. Furthermore, we notice that the acceptable level of automation (i.e., flexibility) $\Lambda$ represents the proportion of the energy consumption that can be reduced, i.e., $\Lambda = 1 - \mathds{1}^\top I_l \overline{u}_l^\ast / \mathds{1}^\top I_l \mathds{1}$.

We propose now a simple rule for tuning the coefficients $\pi_{ui}$ in the utility function \eqref{utility}, which play an important role on the dynamic control scheme we propose in the next section, determining the steady-state solution the controlled system converges to (see \eqref{eq:saved_current}).
Recalling that, without loss of generality, we have decided in Subsection \ref{subsec:C_U} to select the coefficients of the cost function \eqref{cost} such that $\sum_{i=\mathcal{V}} \pi_{ci} = 1$, then, premultiplying both sides of \eqref{minb_sol} by $\mathds{1}^\top$, we obtain
\begin{equation}
\label{eq:Piu}
\mathds{1}^\top \Pi_u \mathds{1} = \frac{\mathds{1}^\top I_l \mathds{1}}{\mathds{1}^\top I_l \overline{u}_l^\ast} -1
= \frac{1}{1-\Lambda}-1,
\end{equation}
implying that the summation of all the coefficients of the utility function \eqref{utility}, i.e., $\sum_{i=1}^n \pi_{ui}$, depends on the acceptable level of automation (i.e., flexibility) $\Lambda$. 
Then, the coefficients $\pi_{ui}$ of the utility function \eqref{utility} have to be selected according to \eqref{eq:Piu}. 
Furthermore, the relation in \eqref{eq:saved_current} suggests to select $\pi_{ui}$ very close to \num{0} if prosumer $i$ does not adopt automation (from \eqref{eq:saved_current} it follows that $\overline{u}_{li}^\ast$ will be very close to \num{1}). Generally, if prosumer $i$ adopts automation, then $\pi_{ui}$ can be selected relatively small if prosumer $i$ is  willing to provide little flexibility. Conversely, $\pi_{ui}$ can be selected relatively large if prosumer $i$ is  willing to provide much flexibility.

{\begin{remark}{\bf (Utility function)}.
\label{rm:tuning}
Note that the utility functions in welfare problems are usually related to economic incentives, which however could make a 100\% renewable energy system not financially sustainable. Then, the tuning of the coefficients of an utility function that is not simply based on economic incentives generally requires an hard endeavour. To the best of our knowledge, this is the first attempt in the literature that provides a motives-based interpretation and estimation of these parameters for control design purposes. This unique feature allows to predict and assess the performance of the controlled system with higher accuracy than other control schemes that generally neglect the individuals' preferences and their impact on the control scheme it self.
\end{remark}}
%
%
%
%
\section{Distributed Primal-dual controller}
\label{sec:controller}
In this section we present a basic primal-dual dynamic controller to achieve Objective \ref{obj:1}. Note that, for the sake of exposition and due to the page limitation, we do not include in the following analysis the constraints discussed in Subsection \ref{subsec:inequalities1}, and refer the interested reader to \cite{stegink2017unifying} and the references therein for the theoretical analysis in presence of inequality constraints. 
Consider the psycho-social-physical welfare \eqref{eq:min2}, i.e., Objective \ref{obj:1}, and let $\lambda_a,\lambda_b: \R_+ \to \mathbb{R}^n$ denote the Lagrange multipliers corresponding to the constraints in \eqref{eq:steady_state_solution_reduced}, respectively. Moreover, let $\lambda:=[\lambda_{a}^\top,\lambda_{b}^\top]^\top$ and $x_c:=[{u}^{\ast\top},x_r^{\ast\top},\lambda^\top]^\top \in \mathcal{X}_c \subseteq \mathcal{U}\times\mathcal{X}_r\times\R^{2n}$. The Lagrangian function corresponding to the optimization problem \eqref{eq:min2} is
\begin{align}
\begin{split}
	\ell (x_c):=&-\alpha W\left(u_l^\ast, I_s^\ast \right) +\frac{\beta}{2}\|u_s^\ast\|^2 +\frac{\gamma}{2}\|V^\ast-V_d\|^2\\
	&+\lambda_a^\top (u_s^\ast-R_s{I}_s^\ast - V^\ast)+\lambda_b^\top (- I_l{u}_l^\ast+{I}_{s}^\ast - \mathcal{L} {V^\ast} ).
\end{split}
\end{align}
Consequently, the first order optimality conditions are given by the Karush-Kuhn-Tucker (KKT) conditions, i.e.,
\begin{subequations}\label{KKT}
\begin{align}
	\beta\overline{u}_s^\ast+\overline\lambda_a&={\bf0}\label{kkt1}\\
	-\alpha I_l  {\Pi}_u^{-1} I_l \left(\mathds{1}-\overline{u}_l^\ast\right)-I_l\overline{\lambda}_b&={\bf0}\label{kkt2}\\
	\alpha  {\Pi}_c^{-1}\overline{I}_s^\ast-R_s\overline{\lambda}_a+\overline{\lambda}_b&={\bf0}\label{kkt3}\\
	\gamma (\overline{V}^\ast-V_d)-\overline\lambda_a - \mathcal{L}\overline{\lambda}_b&={\bf0}\label{kkt4}\\
	  \overline{u}_s^\ast-R_s\overline{I}_s^\ast - \overline{V}^\ast&={\bf0}\label{kkt5}\\
	 - I_l{\overline u}_l^\ast+\overline{I}_{s}^\ast - \mathcal{L} {\overline V^\ast}&={\bf0}.\label{kkt6}
\end{align}
\end{subequations}
{Moreover, we notice that the optimization problem \eqref{eq:min2} is {strictly} convex and satisfies the Slater's condition (the feasibility set ${E}_r$ is nonempty), therefore, strong duality holds~\cite{boyd2004convex}. Hence, $\overline{u}_s^\ast, \overline{u}_l^\ast, \overline{I}_s^\ast$ and $ \overline{V}^\ast$ are optimal if and only if there exist $\overline{\lambda}_a$ and $\overline{\lambda}_b$ satisfying \eqref{KKT}.}

Now, under the assumption that each controller can exchange information among its neighbors through a communication network with the same topology as the physical network, we propose the following distributed control scheme, designed using the primal-dual dynamics of the optimization problem~\eqref{eq:min2}:
\begin{subequations}\label{primal-dual}
\begin{align}
	-\tau_{s}\dot{u}_s^\ast&= \beta {u}_s^\ast + \lambda_a -\nu_s \label{primal-duala}\\
	-\tau_{l}\dot{u}_l^\ast&= -\alpha I_l  {\Pi}_u^{-1} I_l \left(\mathds{1}-{u}_l^\ast\right)-I_l{\lambda}_b - \nu_l \label{primal-dualb}\\
	-\tau_{I}\dot{I}_s^\ast&= \alpha  {\Pi}_c^{-1} {I}_s^\ast-R_s {\lambda}_a + {\lambda}_b\label{primal-dualc}\\
	-\tau_{V}\dot{V}^\ast&= \gamma ({V}^\ast-V_d)- \lambda_a -\mathcal{L}\lambda_b \label{primal-duald}\\
	\tau_a\dot{\lambda}_a&=  {u}_s^\ast-R_s {I}_s^\ast-{V}^\ast  \label{primal-dualf}\\
	\tau_b\dot{\lambda}_b&=- I_l{u}_l^\ast + {I}_{s}^\ast - \mathcal{L} {V^\ast} , \label{primal-dualg}
		\end{align}
\end{subequations}
where $\tau_s,\tau_l,\tau_I,\tau_V,\tau_a, \tau_b>0$ are design parameters and $\nu_s, \nu_l:\mathbb{R}_+\to\mathbb{R}^n$ denote the controller input ports, which will be used later to interconnect the controller \eqref{primal-dual} with the plant \eqref{eq:plant}. 
Let $\nu:=[\nu_s^\top, \nu_l^\top]^\top$.
{Then, we define the forced equilibria set  of system \eqref{primal-dual} as follows:
\begin{equation}
	{E}_{c} := \left\{(\overline{x}_c,\overline{\nu}) \in \mathcal{X}_c\times\R^{2n} | \dot{x}_c={\bf 0} \right\}.
\end{equation}
{Moreover, we notice that, for a given $\overline{\nu}$, the steady-state solution to \eqref{primal-dual} is equivalent to the KKT conditions of the optimization problem \eqref{eq:min2} with the additional penalty $-\overline \nu^\top u$ (see Remark~\ref{rm:penalty} for the physical interpretation of such a penalty). Furthermore, the cost function in \eqref{eq:min2} is strictly convex and, consequently, $\overline{x}_c$ is unique.
}
Now, we establish a passivity property of the controller \eqref{primal-dual} that will be useful later in this section for ensuring the stability of the closed-loop system.
\begin{proposition}{\bf (Passivity property of \eqref{primal-dual})}\label{prop:primal-dual}
	Let $y_c:=[\dot{u}_s^{\ast\top}, \dot{u}_l^{\ast\top}]^\top$ and $\nu_d := [\nu_{sd}^\top, \nu_{ld}^\top]^\top$, $\nu_{sd},\nu_{ld}:\mathbb{R}_+\to \mathbb{R}^{n}$. The following statements hold:
	\begin{itemize}
	\item[(a)] The primal-dual controller \eqref{primal-dual} with $\dot{\nu}=\nu_d$ is passive with respect to the supply rate $\nu_d^\top y_c$ and storage function 
	\begin{equation}
	\label{eq:Sc}
	S_c(x_c,\nu)=\frac{1}{2}\dot{x}_c^\top \tau \dot{x}_c,
	\end{equation}
	with $\tau := \mathrm{blockdiag}\{\tau_s,\tau_l,\tau_I,\tau_V,\tau_v,\tau_a, \tau_b\}$.
	\item[(b)] Let $\nu_d={\bf0}$. The primal-dual controller \eqref{primal-dual} converges to the equilibrium point $(\overline{x}_c,\overline{\nu}) \in E_c$.
	\end{itemize}
\end{proposition}
\begin{IEEEproof}
The storage function $S_c$ in \eqref{eq:Sc} satisfies
\begin{align}
\label{eq:dSc}
\begin{split}
\dot{S}_c &= -\beta \dot{u}_s^{\ast\top}\dot{u}_s^{\ast} - \alpha\dot{u}_l^{\ast\top}I_l {\Pi}_u^{-1}I_l\dot{u}_l^{\ast} -\alpha \dot{I}_s^{\ast\top}  {\Pi}^{-1}_c \dot{I}_s^{\ast}\\
&\hspace{.39cm} - \gamma\dot{V}^{\ast\top}\dot{V}^\ast+{\nu}_{sd}^\top \dot{u}_s^\ast+{\nu}_{ld}^\top \dot{u}_l^\ast\\
&\leq {\nu}_{sd}^\top \dot{u}_s^\ast+{\nu}_{ld}^\top \dot{u}_l^\ast,
\end{split}
\end{align}
along the solutions to \eqref{primal-dual}, concluding the proof of part (a).
For part (b), we conclude from \eqref{eq:dSc} that there exists a forward invariant set $\Omega$ and by LaSalle's invariance principle the solutions that start in $\Omega$ converge to the largest invariant set contained in
\begin{equation}
\Omega \cap \left\{({x}_c,{\nu}) \in \mathcal{X}_c\times\R^{2n}|\dot{u}^\ast={\bf 0}, \dot{I}_s^\ast={\bf0},\dot{V}^\ast={\bf0}, \dot{\nu}={\bf0} \right\}.
\end{equation}
Moreover, from \eqref{primal-duala} and \eqref{primal-dualb} it follows that $\lambda_a$ and $\lambda_b$ are also constant vectors in $\Omega$. 
Then, the solutions that start in $\Omega$ converge to the largest invariant set contained in $\Omega \cap {E}_c$, concluding the proof of part (b).
\end{IEEEproof}

Now, the passive dynamic controller \eqref{primal-dual} is interconnected to the physical  grid \eqref{eq:plant} by choosing $u=u^\ast$, $\nu_s=-I_s$ and $\nu_l=I_lV$. Consequently, we
 obtain the following closed-loop system:
\begin{subequations}
\label{eq:plant_clp_1}
\begin{align}
L_s\dot{I}_{s} & = - R_sI_s - V + u_s^\ast \label{cl1}\\
L\dot{I} & =  -R I -\mathcal{B}^\top V \label{cl2}\\
C\dot{V} & = I_{s} + \mathcal{B}I - I_lu_l^\ast \label{cl3}\\
-\tau_{s}\dot{u}_s^\ast&= I_s + \beta {u}_s^\ast + \lambda_a \label{cl4}\\
	-\tau_{l}\dot{u}_l^\ast&= - I_lV -\alpha I_l  {\Pi}_u^{-1} I_l \left(\mathds{1}-{u}_l^\ast\right)-I_l{\lambda}_b \label{cl5}\\
	-\tau_{I}\dot{I}_s^\ast&= \alpha  {\Pi}_c^{-1} {I}_s^\ast-R_s {\lambda}_a + {\lambda}_b \label{cl6}\\
	-\tau_{V}\dot{V}^\ast&= \gamma\left({V}^\ast-V_d\right)- \lambda_a -\mathcal{L}\lambda_b \label{cl7}\\
	\tau_a\dot{\lambda}_a&= {u}_s^\ast-R_s {I}_s^\ast-{V}^\ast \label{cl9}\\
	\tau_b\dot{\lambda}_b&=- I_l{u}_l^\ast + {I}_{s}^\ast - \mathcal{L} {V^\ast} \label{cl10}. 
\end{align}
\end{subequations}
The set of all feasible operating points of \eqref{eq:plant_clp_1} is defined as
\begin{align}\label{eq:Ecl}
\begin{split}
	{E}_{cl}:=\{&\left(\overline{x},\overline{x}_c\right)\in\mathcal{X}\times\mathcal{X}_c|\\
	&\left(\overline{u}^\ast,\overline{x}\right)\in {E},\left(\overline{x}_c,-\overline{I}_s,I_l\overline{V}\right)\in {E}_{c}\}.
\end{split}
\end{align}

\begin{remark}{\bf (A perhaps surprising additional penalty)}.
\label{rm:penalty}
	Recall that the steady-state conditions of \eqref{primal-dual} represent the KKT conditions of the optimization problem \eqref{eq:min2} with the additional penalty $-\overline{\nu}_s^\top u_s^\ast -\overline{\nu}_l^\top u_l^\ast$, which becomes $\overline{I}_s^\top u_s -\overline{V}I_l^\top u_l$ after interconnecting \eqref{primal-dual} with \eqref{eq:plant}. Then, at the steady-state, from \eqref{eq:steady_state_solution2} we obtain the following expression for the additional penalty:
	\begin{equation}
\overline{I}_s^\top \overline{u}_s -\overline{V}I_l^\top \overline{u}_l = \overline{I}_s^\top R_s \overline{I}_s + \overline{V}^\top \mathcal{B}R^{-1}\mathcal{B}^\top \overline{V},
	\end{equation}
	which implies that also the total power losses in the filters and transmission lines are penalized.
\end{remark}

We can now present the main result of the paper, i.e., the closed-loop stability.
\begin{proposition}{\bf (Stability)}.\label{prop:closed-loop_sys}
Let Assumption \ref{ass:desired_voltages} hold and assume $E_{cl}$ in \eqref{eq:Ecl} to be nonempty.
The closed-loop system \eqref{eq:plant_clp_1} stabilizes to the operating point $(\overline{x},\overline{x}_c) \in E_{cl}$.
\end{proposition}
\begin{IEEEproof}
The storage function
\begin{equation}
S_{cl}(x,x_c) := S(u^\ast,x) + S_c(x_c,-I_s,I_lV)
\end{equation}
satisfies 
\begin{align}
\begin{split}
\dot{S}_{cl}= &-\dot{I}_s^\top R_s \dot{I}_s -\dot{I}^\top R \dot{I} -\beta \dot{u}_s^{\ast\top}\dot{u}_s^{\ast}\\
& - \alpha\dot{u}_l^{\ast\top}I_l {\Pi}_u^{-1}I_l\dot{u}_l^{\ast} -\alpha \dot{I}_s^{\ast\top}  {\Pi}_c^{-1} \dot{I}_s^{\ast} - \gamma\dot{V}^{\ast\top}\dot{V}^\ast,
\end{split}
\end{align}
along the solutions to the closed-loop system \eqref{eq:plant_clp_1}. 
Therefore, there exists a forward invariant set $\Omega$ and by LaSalle's invariance principle the solutions that start in $\Omega$ converge to the largest invariant set contained in
\begin{align}
\begin{split}
\Omega \cap \{(x,{x}_c) \in \mathcal{X}\times\mathcal{X}_c|&\dot{I}_s={\bf0}, \dot{I}={\bf0},\\
& \dot{u}^\ast={\bf 0}, \dot{I}_s^\ast={\bf0},\dot{V}^\ast={\bf0}, \dot{\nu}={\bf0} \}.
\end{split}
\end{align}
Then, from the proofs of Propositions \ref{prop:microgrid} and \ref{prop:primal-dual}, we conclude that the solutions starting in $\Omega$ converge to the largest invariant set contained in $\Omega \, \cap \, {E}_{cl}$.  
Moreover, observing from \eqref{eq:unique_ss} that $\overline x$ is uniquely determined by $\overline u = \overline{u}^\ast$, we can conclude from \eqref{cl9} and \eqref{cl10} that $\overline{I}_s = \overline{I}_s^\ast$, $\overline{V} = \overline{V}^\ast$ and $\overline{I} = -R^{-1}\mathcal{B}^\top \overline{V}^\ast$, i.e., the physical state variables coincide with the corresponding optimization variables. 
\end{IEEEproof}

%
%
%
%
\section{Simulation results}
\label{sec:results}

\begin{figure}
\centering
\begin{small}
\begin{tikzpicture}[>=stealth',shorten >=1pt,auto,node distance=1.85cm,
                    semithick]
  \tikzstyle{every state}=[circle,thick,draw=black,fill=black!3,text=black,minimum size=1.0cm]
  \node[state] (A)                    {Pr 1};
  \node[state]         (B) [right of=A] {Pr 2};
  \node[state]         (C) [right of=B] {Pr 3};
  \node[state]         (D) [right of=C] {Pr 4};
  \node[state]         (E) [right of=D] {Pr 5};
  
   \node[state]         (F) [below of=E] {Pr 6};
   \node[state]         (G) [left of=F] {Pr 7};
   \node[state]         (H) [left of=G] {Pr 8};
   \node[state]         (I) [left of=H] {Pr 9};
   \node[state]         (L) [left of=I] {Pr 10};

  \path[->] 
  (A) edge			node {$I_{1,2}$} (B)
  (A) edge	[left]	     	node {$I_{1,10}$} (L)
  (B) edge              	node {$I_{2,3}$} (C)
  (C) edge              	node {$I_{3,4}$} (D)
  (D) edge              	node {$I_{4,5}$} (E)
  (E) edge			node {$I_{5,6}$} (F)
  (F) edge			node {$I_{6,7}$} (G)
  (G) edge			node {$I_{7,8}$} (H)
  (H) edge			node {$I_{8,9}$} (I)
  (I) edge			node {$I_{9,10}$} (L);

  \path[<->] 
  (A) edge [bend right, dashed, blue]          	node {} (B)
  (A) edge	 [bend left, dashed, blue]	     		node {} (L)
  (B) edge [bend right, dashed, blue]          	node {} (C)
  (C) edge [bend right, dashed, blue]          	node {} (D)
  (D) edge [bend right, dashed, blue]          	node {} (E)
  (E) edge [bend right, dashed, blue]          	node {} (F)
  (F) edge [bend right, dashed, blue]          		node {} (G)
  (G) edge [bend right, dashed, blue]          	node {} (H)
  (H) edge [bend right, dashed, blue]          	node {} (I)
  (I) edge [bend right, dashed, blue]          		node {} (L)
  ;
\end{tikzpicture}
\caption{Scheme of the considered smart grid with 10 prosumers (Pr). The solid arrows indicate the positive direction of the current flows through the power network, while the dashed lines represent the communication network.}
\label{fig:microgrid_example}
\end{small}
\end{figure}
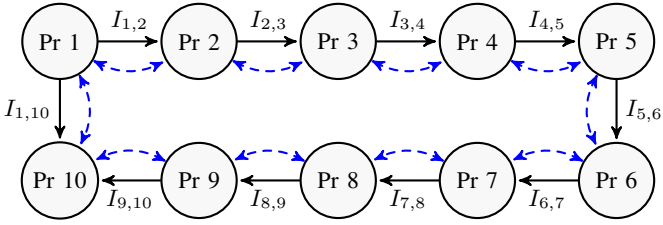

In this section, the proposed distributed optimal primal-dual controller \eqref{primal-dual} is assessed in simulation, by implementing a smart grid comprising 10 prosumers (Pr) connected as illustrated in Figure \ref{fig:microgrid_example}.  

\begin{table}[t]
	\caption{Physical Parameters}
	\centering
	{\begin{tabular}{lc | l}			
											&	&Interval\\			
			\hline
			& & \\
			$R_{si}$	&(\si{\milli\ohm})	&$[1.0,2.0]$		\\
			$L_{si}$	&(\si{\milli\henry})	&$[1.8,3.0]$		\\
			$C_{si}$	&(\si{\milli\farad})	&$[1.7,2.5]$		\\
			$I_{li}$ 	&(\si{\ampere})		&$[6.0,14.0]$	\\
			$R_{k}$	&(\si{\milli\ohm})	&$[50,100]$		\\
			$L_{k}$	&(\si{\micro\henry})	&$[2.0,3.0]$		
	\end{tabular}}
	\label{tab:parameters1}
\end{table}


For each node and line of the smart grid we select a random value in the intervals reported in Table~\ref{tab:parameters1}. For all the prosumers  $i \in \mathcal{V}$, we select $V_{di}=$ \SI{380}{\volt} and, in \eqref{eq:us_limits}, the minimum and maximum permitted values of the voltage $V_{i}$ are selected as $V_i^{\min}=$ \SI{379.3}{\volt} and $V_i^{\max}=$ \SI{380.7}{\volt}. 
All the $\tau$-parameters of the proposed primal-dual controller \eqref{primal-dual} are selected equal to \num{1}, and in the optimization problem \eqref{eq:min2} we choose $\alpha=$ \num{1e6}, $\beta=$ \num{1e-6} and $\gamma=$ \num{1}. For the sake of clarity of exposition, we select the coefficients $\pi_{ci}$ in the cost function \eqref{cost} equal each other for all the prosumers $i \in \mathcal V$, i.e., $\Pi_c = 0.1\,\mathds{I}$. Then, from \eqref{mina_sol} we expect that all the prosumers equally share the total demand.

Now, we estimate based on data from Eurostat\footnote{{\url{https://ec.europa.eu/eurostat/statistics-explained/index.php/Energy_consumption_in_households#Energy_consumption_in_households_by_type_of_end-use}}} what would be the maximum proportion of flexibility within an average household from a technical perspective, representing the proportion of electricity consumed within a household that could potentially be automated. We estimate this number to lay around \num{50}\%, reflecting the energy use for heating/cooling (accounting for roughly \num{60}\% of the total household energy use, of which we assume 2/3 could be automated) and electrical appliances (accounting of roughly \num{20}\% of the total household energy use, of which we assume halve could be automated).
This percentage gives an estimation of what levels of flexibility would be technically possible, i.e., we select $\Psi$ in \eqref{eq:Lambda} equal to \num{0.5}. Thus, if prosumer $i$ fully adopts automation, then $u_{li}^{\min} = 1 - \Psi =$ \num{0.5}, otherwise, if prosumer $i$ does not adopt automation, then $u_{li}^{\min} =$ \num{1}.

Furthermore, as explained in Subsection \ref{subsec:social}, not all prosumers who adopt automation will accept flexibility to the same degree, leading to different values of the coefficients $\pi_{ui}$ appearing in the utility function \eqref{utility} of prosumer $i \in \mathcal{V}$.
\begin{table}[t]
	\caption{Psycho-social-physical Parameters}
	\centering
	{\begin{tabular}{c| l| c| c| c| c}			
			$j$	&Appliance	&$\mu_j$	&$\theta_j$	&$\epsilon_j$	&$\omega_j$	\\			
			\hline
			& & & & &\\
			1	&refrigerator	&\num{0.548}	&\num{0.066}	&\num{0.049}	&\num{0.05}\\
			2	&fridge	&\num{0.532}	&\num{0.070}	&\num{0.044}	&\num{0.05}\\
			3	&dish washer	&\num{0.614}	&\num{0.077}	&\num{0.028}	&\num{0.01}\\
			4	&washing machine	&\num{0.664}	&\num{0.074}	&\num{0.033}	&\num{0.02}\\
			5	&tumble dryer	&\num{0.607}	&\num{0.071}	&\num{0.026}	&\num{0.02}\\
			6	&thermostat	&\num{0.624}	&\num{0.071}	&\num{0.039}	&\num{0.85}\\
			
	\end{tabular}}
	\label{tab:parameters2}
\end{table}
To estimate the acceptable level of automation within a given local energy grid, we performed a questionnaire study among 830 Dutch citizens (of which 794 had valid data) in which we, amongst others, measured individuals' likelihood to adopt automation for different appliances (transformed to range from 0 (definitely not adopting flexibility) to 1 (definitely adopting flexibility)), self-transcendence values (STV) and self-enhancement values (SEV) (both standardized so that the mean equals 0 and the standard deviation equals 1). A detailed description of the methods and scales used can be found in the Appendix.   
From these data, we calculated the average likelihood of adoption of automated control of the appliance $j\in\mathcal{A}:=\{1,\dots,6\}$, denoted in Subsection \ref{subsec:social} by $\mu_j$ (see Table \ref{tab:parameters2}). 
Moreover, to make estimations about specific subpopulations, we generated a linear regression model to predict the likelihood of adoption of automation of the appliance $j\in\mathcal{A}$ based on the endorsement of STV and SEV. The likelihood of adoption of automation of the appliance $j\in\mathcal{A}$ is given by \eqref{eq:rho_j}, where the regression coefficients are reported in Table \ref{tab:parameters2}.
Then, the acceptable level of automation (i.e., flexibility) $\Lambda$ can be computed as in \eqref{eq:Lambda}, where the weights $w_j, j\in\mathcal{A}$ are reported in Table \ref{tab:parameters2} and have been selected based on the common wattage of the appliance $j$ and an estimate of its average use in an typical household. 

We now investigate four different scenarios. In the first one we let the proposed distributed optimal primal-dual controller~\eqref{primal-dual} use the maximum amount of flexibility that is \lq technically' feasible, i.e., $\Lambda=\Psi$. In the second scenario instead, using \eqref{eq:Lambda}, we consider an amount of flexibility that is acceptable to an average subpopulation of our sample (i.e., scoring average (0) on STV and SEV). In the third scenario we suppose that the considered energy community is strongly pro-social/environmental (i.e., high STV) and accordingly less egoistic/hedonic (i.e., low SEV). In the last scenario, we suppose that the considered energy community is strongly egoistic/hedonic (i.e., high SEV) and accordingly less pro-social/environmental (i.e., low STV).
In all the scenarios, we assume that all the prosumers of the considered energy community adopt automation. Then, we select $u_{li}^{\min} =$ \num{0.5}, for $i=1,\dots,10$. Also, for the coefficients $\pi_{ui}, i=1,\dots,10$, we select normally distributed values, using the mean and standard deviation for the SEV (see the Appendix). 

{\bf Scenario 1}. Let the proposed controller~\eqref{primal-dual} use the maximum amount of flexibility that is \lq technically' feasible, i.e., $\Lambda=\Psi$.
We can observe in Figure \ref{fig:s1} that the voltage at the PCC of each prosumer is stable and remains within the bounds (dashed lines) that ensure a proper functioning of the connected appliances and a safe and reliable functioning of the overall grid. Indeed, the average voltage is \SI{380.05}{\volt}, which is very close to the voltage reference. Moreover, we notice that also the currents are stable and current sharing is achieved. 
Furthermore, we observe that also $u_l$ is stable and within the bounds (dashed line). Finally the total consumption reduction is $\mathds{1}^\top I_l(\mathds{1}-\overline{u}_l) =$ \SI{39.7}{\ampere} (44.1\%).

{\bf Scenario 2}. Let the proposed controller~\eqref{primal-dual} use the flexibility that is  \lq socially' feasible in an \lq average' energy community, i.e., we select in \eqref{eq:rho_j} $\mathrm{STV} = \mathrm{SEV} =$ \num{0} and compute $\Lambda$ as in \eqref{eq:Lambda}.
We can observe in Figure \ref{fig:s2} that the voltage at the PCC of each prosumer is stable and remains within the bounds (dashed lines), with an average value equal to \SI{380.07}{\volt}. Moreover, we notice that also the currents are stable. Yet, in order to comply with the voltage constraints, (equal) current sharing is not perfectly achieved. Furthermore, we observe that also $u_l$ is stable and within the bounds (dashed line). Finally the total consumption reduction is $\mathds{1}^\top I_l(\mathds{1}-\overline{u}_l) =$ \SI{27.3}{\ampere} (30.3\%), which is less than the one in Scenario 1 because prosumers do not accept to provide the maximum amount of flexibility that is \lq technically' feasible.

{\bf Scenario 3}. Let suppose that the considered energy community is strongly pro-social/environmental and less egoistic/hedonic, i.e., we select in \eqref{eq:rho_j} $\mathrm{STV} =$ \num{2} and $\mathrm{SEV} =$ \num{-1} and compute $\Lambda$ as in \eqref{eq:Lambda}.
We can observe in Figure \ref{fig:s3} that the voltage at the PCC of each prosumer is stable and remains within the bounds (dashed lines), with an average value equal to \SI{380.07}{\volt}. Moreover, we notice that also the currents are stable. Yet, in order to comply with the voltage constraints, (equal) current sharing is not perfectly achieved. Furthermore, we observe that also $u_l$ is stable and within the bounds (dashed line). Finally the total consumption reduction is $\mathds{1}^\top I_l(\mathds{1}-\overline{u}_l) =$ \SI{31.3}{\ampere} (34.8\%), which is less than the one in Scenario 1 because prosumers do not accept to provide the maximum amount of flexibility that is \lq technically' feasible. Yet, it is greater than the one in Scenario 2 because flexibility particularly appeals to individuals with stronger STV, likely because of flexibility's environmental benefits, making prosumers in this community more likely to adopt flexibility.

{\bf Scenario 4}. Let suppose that the considered energy community is strongly egoistic/hedonic and less pro-social/environmental, i.e., we select in \eqref{eq:rho_j} $\mathrm{STV} =$ \num{-1} and $\mathrm{SEV} =$ \num{2} and compute $\Lambda$ as in \eqref{eq:Lambda}.
We can observe in Figure \ref{fig:s4} that the voltage at the PCC of each prosumer is stable and remains within the bounds (dashed lines), with an average value equal to \SI{380.07}{\volt}. Moreover, we notice that also the currents are stable. Yet, in order to comply with the voltage constraints, (equal) current sharing is not perfectly achieved. Furthermore, we observe that also $u_l$ is stable and within the bounds (dashed line). Finally the total consumption reduction is $\mathds{1}^\top I_l(\mathds{1}-\overline{u}_l) =$ \SI{27.6}{\ampere} (30.6\%), which is less than the one in Scenario 1 because prosumers do not accept to provide the maximum amount of flexibility that is \lq technically' feasible. Yet, it is almost equal to  the one in Scenario 2 and definitely less than the one in Scenario 3. This can be explained by the notion that individuals see flexibility to primarily benefit STV (e.g., pro-environmental), more so than SEV (e.g., financial) (see Table \ref{tab:parameters2}), making flexibility less attractive to prosumers who care much about SEV, but only little about STV.

\begin{figure}
	\includegraphics[width=\columnwidth]{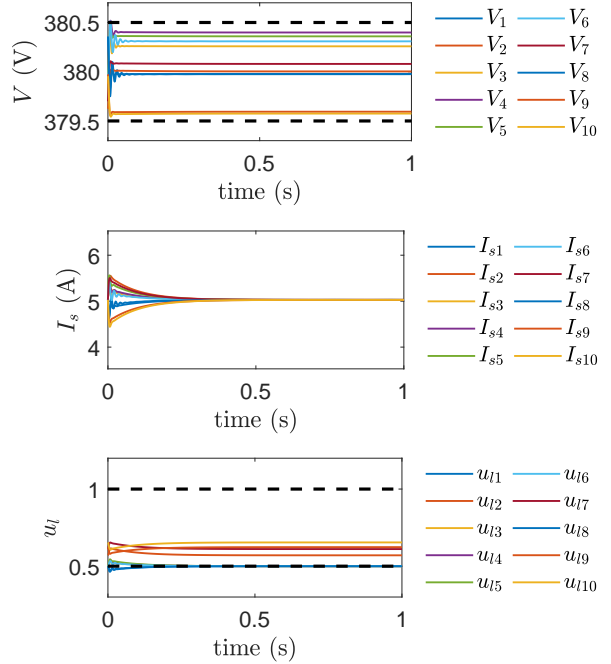}
	\caption{Scenario 1. From the top: voltage at the PCC; generated current; load control input.}\label{fig:s1}
\end{figure}

\begin{figure}
\vspace{-.55cm}
	\includegraphics[width=\columnwidth]{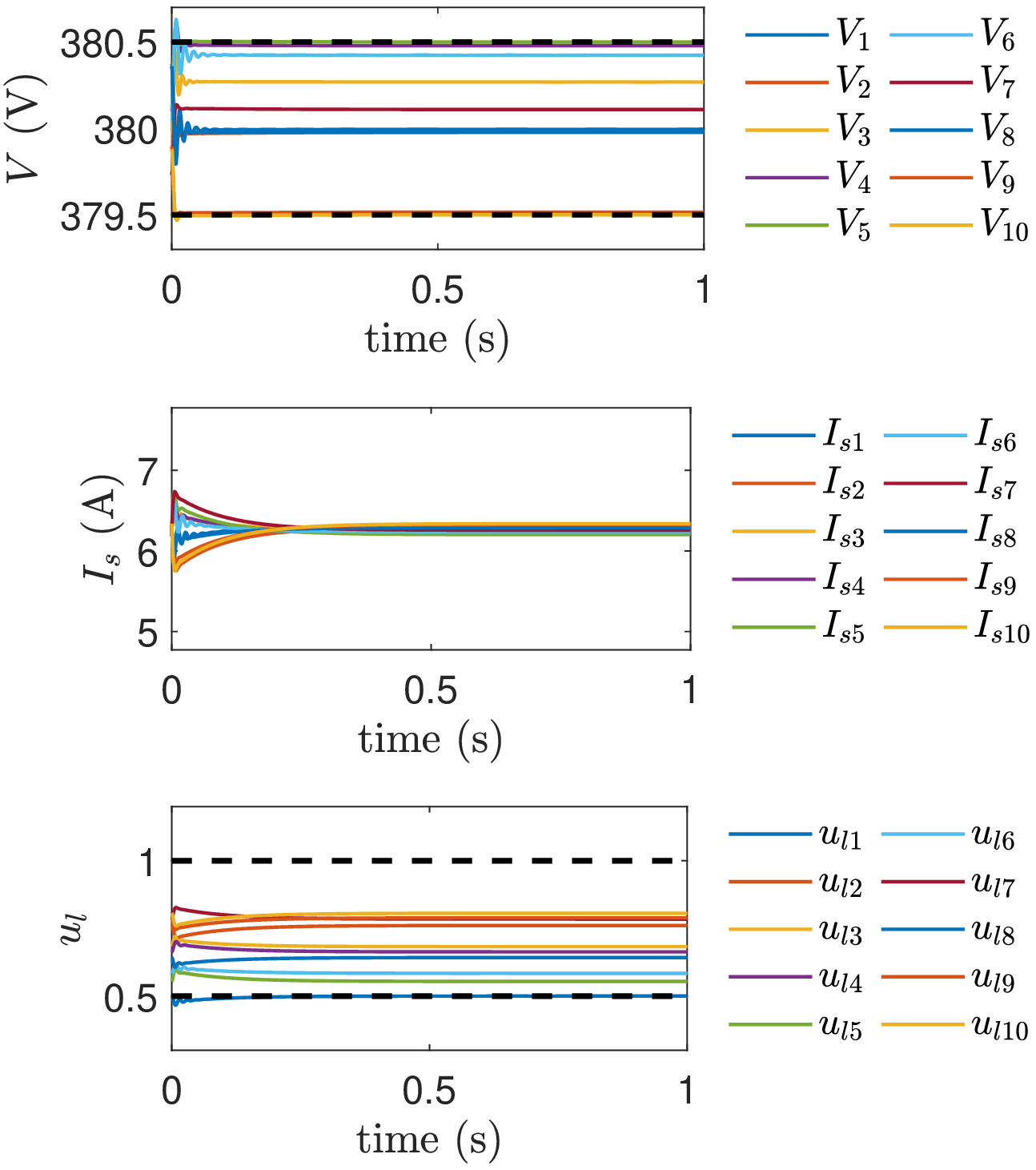}
	\caption{Scenario 2. From the top: voltage at the PCC; generated current; load control input.}\label{fig:s2}
\end{figure}

\begin{figure}
	\includegraphics[width=\columnwidth]{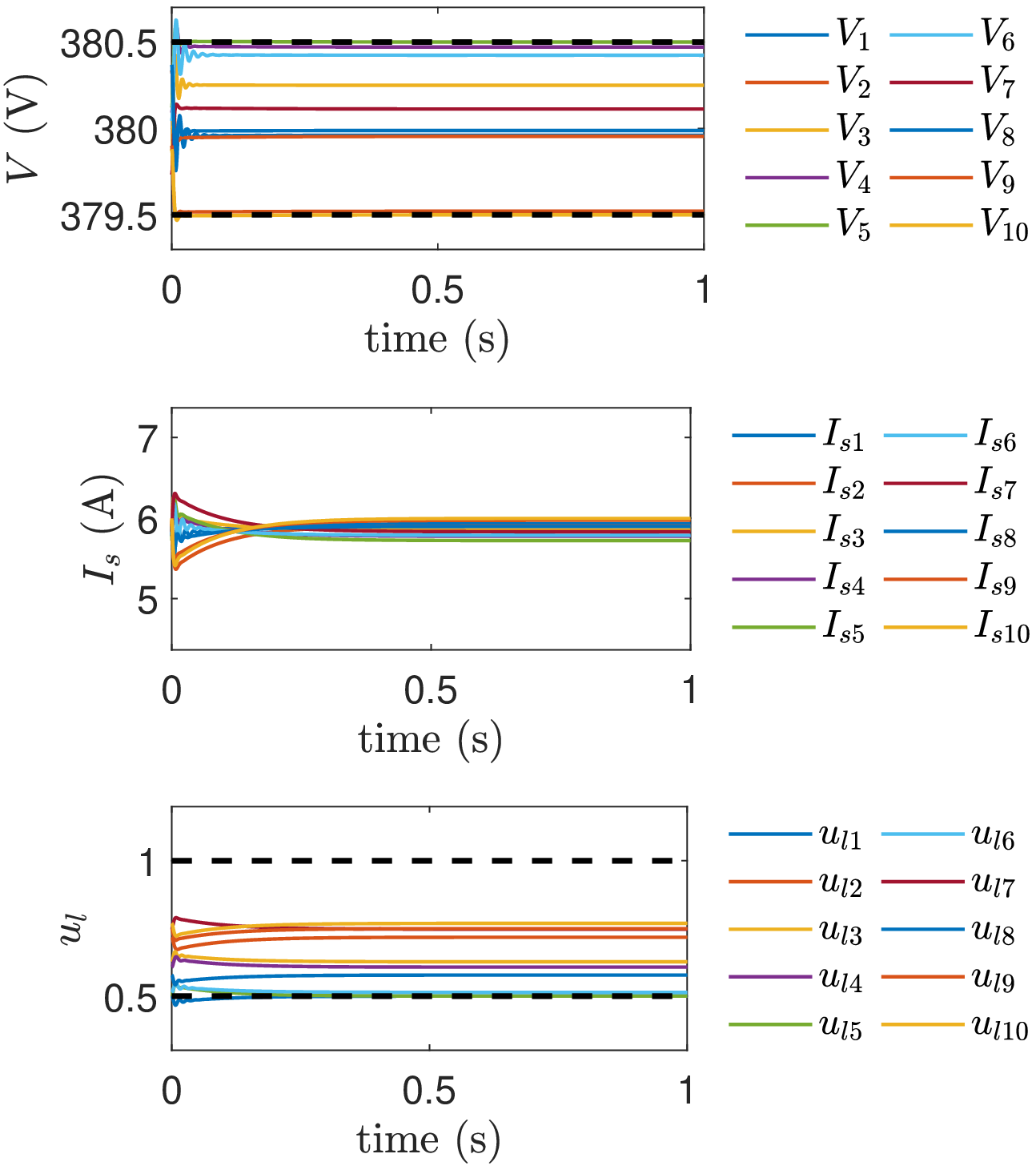}
	\caption{Scenario 3. From the top: voltage at the PCC; generated current; load control input.}\label{fig:s3}
\end{figure}

\begin{figure}
	\includegraphics[width=\columnwidth]{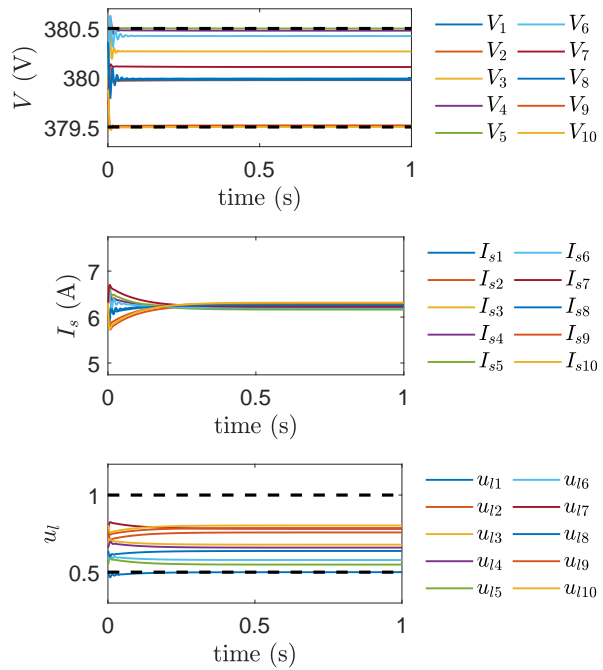}
	\caption{Scenario 4. From the top: voltage at the PCC; generated current; load control input.}\label{fig:s4}
\end{figure}

\begin{remark}{\bf (Different data)}.
\label{rm:different_data}
The values of $\mu_j$, $\theta_j$ and $\epsilon_j$ reported in Table \ref{tab:parameters2} are based on data from a Dutch sample, which might not represent populations of other countries, nor future populations. Other populations might for instance be more or less familiar with automated control, and therefore be more or less likely to adopt such technologies. Importantly, we demonstrate how psycho-social factors could be integrated in the design of control schemes for energy systems, and provide a usable framework for how this could be achieved. Yet, the exact coefficients and numbers could be updated and adjusted based on new insights and data from targeted populations as well as societal and system changes.
\end{remark}

%
%
%
%
\section{Conclusions and Future Research}
\label{sec:conclusions}
Our work shows how incorporating key motives of prosumers, here represented by self-transcendence and self-enhancement values, in a DC smart grid can affect the outcomes and performance of the proposed primal-dual controller. 
Our results show a big discrepancy between the technical potential of a control scheme (i.e., what is possible if flexibility is fully adopted) and what is psycho-socially feasible (i.e., what is possible when considering user preferences and motives). Our data suggests that far less flexibility is adopted than would technically be possible, which limits the impacts control schemes can have on the energy system. Yet, the level of adoption of flexibility, and thus the effectiveness of our control schemes, seems to vary among communities. Specifically, flexibility seems to be primarily perceived as benefiting self-transcendence values (e.g., having environmental benefits), more so than benefiting self-enhancement values (e.g., having financial benefits), which makes communities with strong self-transcendence values more likely to adopt flexibility, and thus to benefit from our proposed control scheme.
Better understanding and incorporating psycho-social factors in models could facilitate the effective implementation of proposed solutions, enabling decision makers to select those communities in which specific solutions will yield the most positive outcomes. Moreover, our outcomes may assist in effectively promoting solutions, highlighting to which values one could appeal, for instance because these values are strong predictors of adoption and flexibility, or because these values are strongly endorsed within a population, which both appear to be the case for self-transcending values within our studied population (i.e., Dutch citizens). Thus, by including psycho-social factors to models, physical and technical as well as psycho-social constraints can be taken into account, which enhances the feasibility of policies.

That said, it is also important to note that our efforts are a first and simplified demonstration of how psycho-social factors could be incorporated in the modelling of energy systems that can be improved and extended by future research. Specifically, future research could identify additional psycho-social factors that affect the uptake and use of automation and integrate them in our, or other, control schemes.  The proposed regression models that contained self-transcendence and self-enhancement values as predictors were able to explain around 10\% of the variance in adoption of automated control, which is reasonable for social sciences, but also leaves much room for further improvements. For example, various factors from psychology (e.g., personal and social norms), but also from other disciplines (e.g., economics, law), could bring key advancements through our, and other, energy system models. In addition, future research could aim at measuring the focal variables, in particular adoption of flexibility, in more accurate ways. Here, we estimated the likelihood of adoption based on self-reported intentions, which may not accurately reflect actual adoption rates. Hence, our activities were mainly aimed at $i$) highlighting the relevance of incorporating psycho-social factors in energy models and $ii$) providing means to do so as to stimulate and facilitate future research on such integrations, rather than providing ready-to-use control schemes.

The latter issue also reflects a broader issue that should be addressed to enable the better integration of psycho-social factors in energy system models. That is, in many social sciences energy behaviours are operationalized as self-reported intentions, likelihoods or frequencies of a behaviour, which are difficult to translate into more objective, physical numbers that could be incorporated in optimal control schemes (e.g., proportion of people adopting automated control, energy used, etc.). Future social science studies could employ more objective, physical measures as to facilitate the integration of psycho-social factors into optimal control schemes. In addition, the considered optimization problem is relatively static in that it focuses on psycho-social factors at one moment in time. However, prosumers interact with each other, and will react on each other, which would make it interesting to research the dynamics (e.g. arising from studies on opinion dynamics and social networks) of psycho-social aspects over time, and how such developments will impact optimal control schemes. 
Working on these aspects, for which our paper provides initial solutions, could greatly contribute to a transition towards 100\% renewable energy systems, making solution more acceptable, desirable and realistic.

\appendix[Supplementary methods]

\subsection{Respondents} 
The psycho-social data presented in the manuscript are based on a questionnaire study conducted among 830 Dutch citizens. Data was collected via Qualtrics, an online research platform who approached a random selection of their Dutch panel. After completion of the questionnaire, respondents were rewarded with credits, which they could use for online purchases. 
Out of the total 830 respondents, 794 passed the quality check. Those respondents who did not pass the quality check, and thus appeared to not have filled out the questionnaire seriously (e.g. selecting the same response option for every question), were removed from the analyses. The final dataset consisted of 409 women and 383 men (2 did not indicate their gender), with a mean age of 42.2 years old (SD = 14.64).

\subsection{Procedure}
Before starting the questionnaire, respondents were informed about the study, the data collection and storage procedures, and their rights as a respondent (following standard procedures of psychological research in the Netherlands), and asked for their consent. Those respondents who consent with participating in the study were presented with the questionnaire, which – amongst others – measured the likelihood that they would let automation control several electrical appliances in their house, as well as their self-transcendent and self-enhancement values. At the end of the questionnaire we also asked respondents about more socio-demographic variables; specifically, gender, age, education level, income and household composition. The inclusion of those variables as covariates in our model hardly changed our outcomes, nor did they clearly contribute to the prediction of adoption of automation, which is why we did not report these variables in the manuscript.

\subsection{Measures}

	\subsubsection{Likelihood of adopting automation ($\mu$)} Participants were asked to indicate on a 5-point scale ranging from 1 (not at all) to 5 (certainly yes) to what extent they would be willing to let automation technology control their thermostat (M = 3.50, SD = 1.20), refrigerator (M = 3.19, SD = 1.26), freezer  (M = 3.13, SD = 1.31), dishwasher (M = 3.45, SD = 1.32), washing machine (M = 3.66, SD = 1.16) and dryer (M = 3.43, SD = 1.32). For the analyses presented in the paper, the scores were transformed (i.e., we subtracted 1 from the original score, and divided it by 4) to range from 0.00 (here interpreted as reflecting a 0\% likelihood that the respondent would automate the appliance) to 1.00 (here interpreted as reflecting a 100\% likelihood that the respondent would automate the appliance).\\
	
	\subsubsection{Values (STV and SEV)} Participants' values were measured by asking them to indicate on a 9-point scale, ranging from -1 (opposes my values) to 0 (not important) to 7 (extremely important), how important each of 16 items were to them as a guiding principle in their life \cite{bouman2018measuring}. Self-transcendence values were measured with a scale consisting of 8 items: \lq Equality (equal opportunities for all)', \lq Respecting the earth (harmony with other species)', \lq Unity with nature (fitting into nature)', \lq A world at peace (free of war and conflict)', \lq Social justice (correcting injustice, care for the weak)', \lq Protecting the environment (preserving nature)', \lq Helpful (working for the welfare of others)', and \lq Preventing pollution (protecting natural resources)'. We computed mean scores on these items; the Cronbach's alfa of this scale is .90 (M = 4.80, SD = 1.36). Self-enhancement values were also measured with a scale consisting of 8 items: \lq Power (control over others, dominance)', \lq Pleasure (gratification of desires)', \lq Wealth (material possessions, money)', \lq Authority (the right to lead or command)', \lq Enjoying life (enjoying food, sex, leisure etc.)', \lq Influential (having an impact on people and events', \lq Self-indulgent (doing pleasant things)', and \lq Ambitious (hard-working, aspiring)'. We computed mean scores on these items; the Cronbach's alfa of this scale is .79 (M = 3.22, SD = 1.23). For the analyses presented in the paper, we standardized the computed self-transcendence and self-enhancement values, so that the mean reflects 0 and the standard deviation 1.

\balance
\bibliographystyle{IEEEtran}
\bibliography{reference}

%

%

\end{document}